\documentclass[aps,prl,twocolumn,groupedaddress,showpacs,floatfix,superscriptaddress,longbibliography]{revtex4-2}
\usepackage[plainpages=false,pdfpagelabels,colorlinks=true,linkcolor=red,urlcolor=blue,citecolor=blue,pdftitle={Title},pdfauthor={},pdfdisplaydoctitle=true,pdfduplex=DuplexFlipLongEdge]{hyperref}
\usepackage{siunitx}
\usepackage{mhchem}
\usepackage{epsfig}
\usepackage{graphicx}
\usepackage[utf8]{inputenc}
\usepackage{amsmath,amssymb}
\usepackage[dvipsnames]{xcolor}

\usepackage[normalem]{ulem}

\usepackage{tabularx}
\usepackage{amsmath}

\usepackage[T1]{fontenc}
\usepackage[utf8]{inputenc}  
\usepackage{booktabs}

\begin{document}

\title{Polaron-mediated metal-insulator transition and proton conduction in hydrogenated nickelate perovskites}
\author{Hang Ma}
\affiliation{School of Physics and Astronomy, Beijing Normal University, and Key Laboratory of Multiscale Spin Physics (Beijing Normal University), Ministry of Education, Beijing 100875, China\\}  
\author{Tianxing Ma}
\email{txma@bnu.edu.cn}
\affiliation{School of Physics and Astronomy, Beijing Normal University, and Key Laboratory of Multiscale Spin Physics (Beijing Normal University), Ministry of Education, Beijing 100875, China\\} 
\author{Ying Liang}
\email{liangying@hebtu.edu.cn}
\affiliation{College of Physics, Hebei Normal University, and Hebei Advanced Thin Films Laboratory, Shijiazhuang 050024, China}
\affiliation{School of Physics and Astronomy, Beijing Normal University, and Key Laboratory of Multiscale Spin Physics (Beijing Normal University), Ministry of Education, Beijing 100875, China\\}

\begin{abstract}
Nickel-based perovskites, owing to their spontaneous hydrogen uptake and the dramatic increase in resistivity upon hydrogenation, have emerged as promising candidates for proton-conducting fuel cell electrolytes. However, the mechanism of the hydrogen-induced metal-insulator transition (MIT) in rare-earth nickelates remains under debate, particularly regarding whether the doped electrons occupy Ni e$_g$ states or O 2p ligand hole states. Here, we reveal a comprehensive MIT mechanism using first-principles calculations on NdNiO$_3$: the electrons introduced by hydrogen doping occupy the O 2p ligand hole states of the Ni-O hybridized d$_8$L configuration, promoting electron-polaron formation. The resulting electron polarons, together with proton polarons, weaken the Ni-O hybridization and thereby drive the originally itinerant Ni e$_g$ electrons toward localization. This generates a local d8 (t$_{2g}$$^6$e$_g$$^2$) electronic configuration, leading to a Mott transition. In addition, we also find that compared with NdNiO$_3$, SmNiO$_3$ with a smaller A-site ionic radius more readily absorbs hydrogen but exhibits weaker proton diffusion capability. Hydrogenation promotes proton permeation along the [001] direction via the intraoctahedral transfer, whereas the overall proton diffusivity is reduced. These results provide guidance for experimental screening of strongly correlated oxides as electrolyte materials and offer theoretical insights for enhancing proton conductivity in rare-earth nickelates.
\end{abstract}

\date{Version 16.0 -- \today}

\maketitle


\section{Introduction}
In the context of the global energy transition, solid oxide fuel cells (SOFCs), as one of the representative energy conversion technologies, have attracted considerable attention due to their high efficiency in converting chemical energy into electricity and their environmentally benign byproducts\cite{abe2019hydrogen,XU2022115175,BICER20203670}. Compared to conventional oxygen-ion conducting SOFCs that typically require high operating temperatures above 800~\si{\celsius}\cite{TIMURKUTLUK20161101,doi:10.1021/acsami.3c09025}, proton-conducting SOFCs, which exhibit excellent electrochemical performance at intermediate and low temperatures (400-600~\si{\celsius}), have emerged as a promising alternative\cite{ding2020self}. This enhanced performance is primarily attributed to the high proton conductivity in the electrolyte layer, highlighting the critical importance of developing electrolyte materials with high proton conductivity for improving the overall performance of proton-conducting SOFCs\cite{somekawa2016physicochemical}.

Conventional proton-conducting electrolyte materials are primarily II-IV perovskite oxides, such as BaCeO$_3$, BaZrO$_3$, and SrCeO$_3$\cite{kannan2013chemically,B902343G,FABBRI20101043,Münch01041999}. In these II-IV ABO$_3$-type systems, proton conduction is typically enabled by aliovalent doping at the A- or B-site to introduce oxygen vacancies\cite{zhang2018insight,liu2020microscopic}. Under humid conditions, these vacancies incorporate hydroxyl groups from dissociated water, while the accompanying proton bound to neighboring oxygen ions\cite{domingo2019water}. Proton transport then proceeds via the Grotthuss mechanism\cite{agmon1995grotthuss,kreuer2000complexity}, involving proton rotation at a given oxygen site through lattice-assisted hydrogen-bond breaking, and proton transfer between adjacent oxygen ions along strong hydrogen-bond networks\cite{braun2017experimental,merinov2009proton,KREUER1995157,MUNCH2000183,jing2020role,ma2026hydrogen,samgin2000lattice,du2020cooperative}. Therefore, conventional electrolyte materials require cation substitution to maintain ionic conductivity. In contrast, recently studied strongly correlated oxides can serve as electrolytes in their pure form. The design principle of these materials as electrolytes in proton-conducting fuel cells is as follows: when hydrogen is dissociated into protons and electrons at the anode under the action of a catalyst, the strongly correlated oxide electrolyte undergoes hydrogenation at the three-phase interface and transforms from a metal to an insulator. This process facilitates proton transport through the electrolyte under a chemical potential gradient while suppressing electron flow through the internal circuit to the cathode, thereby driving current through the external circuit\cite{zhou2016strongly}. 

Consequently, the key feature of strongly correlated oxides as electrolytes lies in their spontaneous hydrogen uptake and subsequent MIT transition. Studies have shown that VO$_2$\cite{chen2019gate,zhou2025hydrogen,zhou2022revealing}, SrCoO$_{2.5}$\cite{lu2017electric,wei2025hydrogen,islam2020computational,PhysRevMaterials.3.024603}, and ReNiO$_3$ (Re = La, Pr, Nd, Sm, Eu)\cite{zhou2016strongly,shi2014colossal,chen2020electron,yoo2018metal,lan2020first,gao2023unveiling} all possess hydrogenated phases and undergo electronic-state transitions upon hydrogenation. Among them, ReNiO$_3$ belongs to a series of rare-earth nickelates with the perovskite structure, whose complex electronic phase diagrams and orbital configurations enable applications in the fields of fuel cells and superconductivity\cite{zhou2016strongly,shi2014colossal,chen2020electron,PhysRevB.109.235123,PhysRevLett.109.156402}. Moreover, whereas temperature-induced changes in ReNiO$_3$ only increase the resistivity by three orders of magnitude\cite{medarde1997structural}, hydrogenation can enhance it by as much as eight orders of magnitude\cite{shi2014colossal,gao2023unveiling}. However, the mechanism of the hydrogen-induced MIT transition in ReNiO$_3$ remains under debate. The conventional and prevailing view is that hydrogenation is accompanied by electron filling into the originally fourfold-degenerate Ni orbitals, driving the Ni d-orbital configuration from the itinerant state t$_{2g}$$^6$e$_g$$^1$ to the localized state t$_{2g}$$^6$e$_g$$^2$, thereby introducing intra-orbital Coulomb repulsion and opening a gap between the lower and upper Hubbard bands, resulting in a Mott insulating state\cite{zhou2016strongly,shi2014colossal,yoo2018metal,lan2020first} . In contrast, recent studies on the NdNiO$_3$ system suggest an anti-doping mechanism in which the electrons occupy ligand holes in the O-2p orbital, rather than simply converting Ni$^{3+}$ to Ni$^{2+}$\cite{gao2023unveiling}. Therefore, a deeper understanding of the mechanism underlying the hydrogen-induced MIT transition in ReNiO$_3$ is still highly desirable.

The hydrogen-induced insulating state is necessary for nickelates to serve as proton-conducting SOFC electrolytes. However, their practical performance is dictated by proton transport. SmNiO$_3$, one of the representative ReNiO$_3$ systems, exhibits high ionic conductivity and low proton migration barriers comparable to or even surpassing conventional proton conductors\cite{zhou2016strongly,yoo2018metal,lan2020first,PhysRevB.82.014103}, highlighting its potential for proton transport in low-temperature SOFCs. Although a hydrogen-induced insulating phase has recently been identified in NdNiO$_3$\cite{gao2023unveiling}, its proton transport properties remain largely unexplored. Given that proton diffusion is strongly coupled to lattice distortions\cite{braun2017experimental,merinov2009proton,KREUER1995157,MUNCH2000183,jing2020role,ma2026hydrogen,samgin2000lattice,du2020cooperative}, and that hydrogenation induces significant structural changes in ReNiO$_3$\cite{zhou2016strongly,yoo2018metal,gao2023unveiling}, a systematic investigation of hydrogenation effects on proton transport, as well as the role of A-site rare-earth cations, is therefore warranted. 

In this work, we conduct a comprehensive study of the lattice distortions and the MIT transition induced by hydrogenation, and assess the proton conduction potential of these nickelate perovskite systems. Using first-principles calculations, we find that hydrogen insertion in NdNiO$_3$ elongates the Ni-O bonds. Further analysis reveals a polaron-mediated Mott transition MIT mechanism in which filling of the Ni-O hybridized states induces polaron formation, driving electron localization. In addition, we find that rare-earth nickelates with smaller A-site cations more readily absorb hydrogen spontaneously but exhibit weaker proton diffusivity. Hydrogenation facilitates one-dimensional proton diffusion, while impeding three-dimensional conduction. Our results provide guidance for experimental screening strongly correlated oxides as proton-conducting electrolyte materials.

\begin{table*}[htbp]
\centering
\caption{Lattice constants, bond lengths, bond angles, and structural distortion modes of NdNiO$_3$ before and after hydrogenation, as obtained from our PBEsol+U calculations, experiments\cite{catalano2018rare,middey2016physics}, and previous LDA+U+J results\cite{gao2023unveiling}. Values in parentheses denote the percentage changes upon hydrogenation relative to the pristine system.}
\label{Tab1}
\setlength{\tabcolsep}{8pt}
\renewcommand{\arraystretch}{1.2}
\begin{tabular}{lccccc}
\toprule
& \multicolumn{3}{c}{NdNiO$_3$} & \multicolumn{2}{c}{HNdNiO$_3$} \\
\cmidrule(lr){2-4} \cmidrule(lr){5-6}
& PBEsol+U & LDA+U+J\cite{gao2023unveiling} & Experiment\cite{catalano2018rare,middey2016physics} & PBEsol+U & LDA+U+J\cite{gao2023unveiling}  \\
\midrule
$a$ (\AA) & 5.331 &  & 5.389 & 5.661 (+6.2\%) &   \\
$b$ (\AA) & 5.331 &  & 5.382 & 5.661 (+6.2\%) &  \\
$c$ (\AA) & 7.605 &  & 7.61  & 7.605 &   \\
$\alpha$ ($^\circ$) & 90 &  &  & 90 &   \\
$\beta$ ($^\circ$)  & 90 &  &  & 90 &   \\
$\gamma$ ($^\circ$) & 89.85 &  &  & 85.33 &   \\
\midrule
Ni--O$_{8d}$ (\AA) & 1.925 & 1.945 &  & 2.116 (+9.9\%) & 2.135 (+9.77\%)  \\
Ni--O$_{4c}$ (\AA) & 1.938 & 1.95  &  & 1.993 (+2.8\%) & 2.1 (+7.7\%)   \\
$\angle$Ni--O$_{8d}$--Ni ($^\circ$) & 156.54 & 155.8 &  & 142.14 (-9.2\%) & 145.6 (-6.5\%)   \\
$\angle$Ni--O$_{4c}$--Ni ($^\circ$) & 157.62 & 155.8 &  & 145.15 (-7.91\%) & 144 (-7.57\%)  \\
\midrule
a$^-$a$^-$c$^0$ (\AA/f.u.) & 1.1341 &  &  & 1.7504 &   \\
a$^0$a$^0$c$^+$ (\AA/f.u.) & 0.7301 &  &  & 1.39 &    \\
\bottomrule
\end{tabular}
\end{table*}


\section{Computational Method}
\begin{figure*}[htbp]
\centering
\includegraphics[scale=0.49]{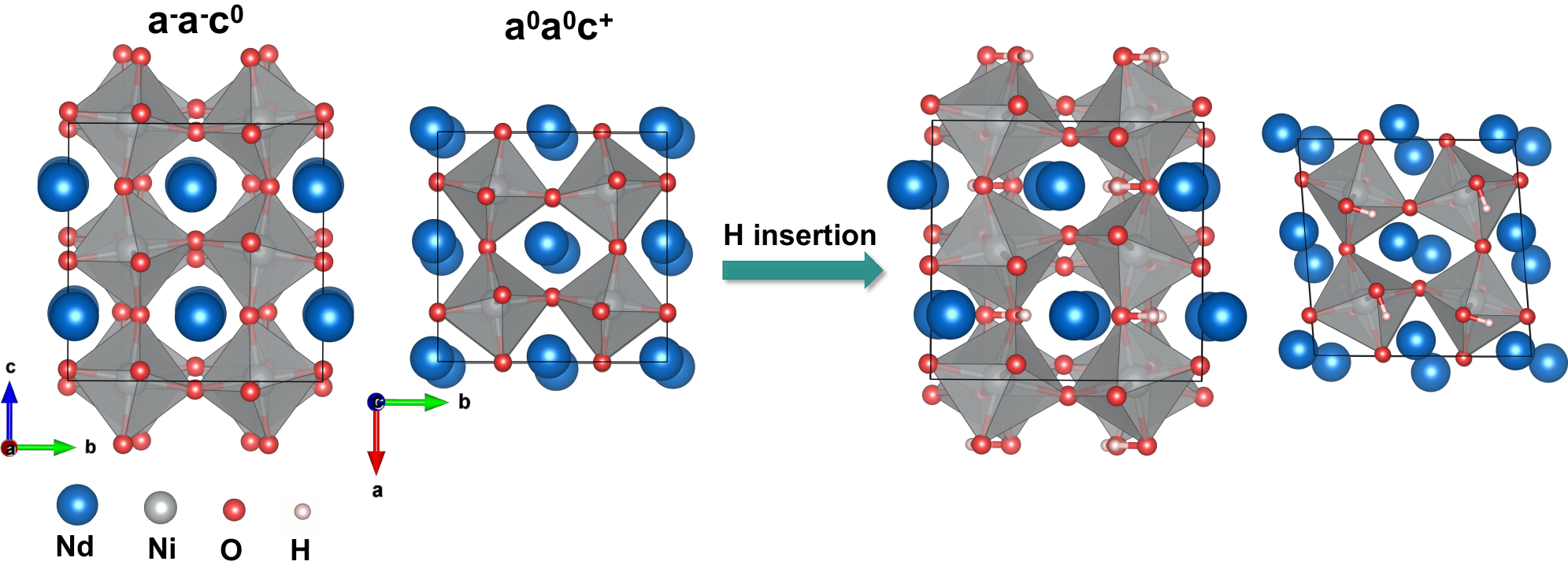}
\caption{Schematic illustration of the octahedral rotation patterns in NdNiO$_3$ and the lattice distortions induced by hydrogenation.}
\label{Fig1}
\end{figure*}

First-principles density functional theory (DFT) calculations were performed using the Vienna ab initio simulation package (VASP)\cite{kresse1996efficiency,kresse1996efficient}. Spin-polarized projector augmented-wave (PAW) pseudopotentials\cite{PhysRevB.59.1758} were employed with the following valence configurations: Nd\_3/Sm\_3 (5s$^2$5p$^6$5d$^1$6s$^2$, with the f electrons frozen in the core\cite{kiejna2006comparison}), Ni (3d$^8$4s$^2$), and O (2s$^2$2p$^4$). The exchange-correlation functional was treated within the generalized gradient approximation (GGA) using the Perdew-Burke-Ernzerhof revised for solids (PBEsol) form\cite{PhysRevLett.100.136406}. Electron correlation effects were described within the rotationally invariant DFT+U approach of Dudarev et al.\cite{PhysRevB.57.1505}, with an effective Hubbard parameter U$_{eff}$=2.0~eV applied to the Ni 3d orbitals, consistent with previous PBEsol+U studies\cite{yoo2018metal,varignon2017complete}. Larger values of U have been shown to yield a monoclinic ferromagnetic ground state that is inconsistent with experimental observations for rare-earth nickelates\cite{varignon2017complete}. The cutoff energy of the plane wave basis was set to 520~eV. The electronic self-consistent field (SCF) convergence criterion was set to 10$^{-6}$eV. Structural optimization was completed using the conjugate gradient algorithm until the Hellmann-Feynman forces on each atoms are less than 0.01~eV/\AA.

Since rare-earth nickelates adopt an orthorhombic phase within the operating temperature range of SOFCs\cite{medarde1997structural,girardot2008raman}, a $\sqrt{2} \times \sqrt{2} \times 1$ supercell containing 40 atoms (8 Nd, 8 Ni, and 24 O) was constructed based on the optimized orthorhombic NdNiO$_3$ unit cell. A $\Gamma$-centered $4\times4\times4$ k-point mesh with a spacing of 0.03~\AA$^{-1}$ was employed. Structural relaxations were performed using Gaussian smearing with a width of 0.1 eV, while the electronic structure calculations were carried out using the tetrahedron method with Blöchl corrections\cite{PhysRevB.49.16223}. Based on the optimized NdNiO$_3$ supercell, we constructed the fully hydrogenated H$_x$NdNiO$_3$ (x = 1) configuration by inserting eight protons at apical oxygen sites, which has been identified as the most stable structure at high hydrogen concentration\cite{yoo2018metal,gao2023unveiling}. Starting from this lowest-energy HNdNiO$_3$ configuration, one hydrogen atom was removed from all possible inequivalent sites, and symmetrically distinct configurations were identified and retained. This procedure was iteratively applied to generate all distinct configurations at lower hydrogen concentrations (from x = 0.875 to x = 0.125). Finally, all inequivalent configurations at each concentration were fully relaxed. This approach to sampling hydrogen configurations follows the methodology reported by Pilsun Yoo et al.\cite{yoo2018metal}.

The lattice distortion modes and their amplitudes of the ground-state structures before and after hydrogenation were extracted using symmetry-mode analysis implemented in the ISOTROPY applications\cite{stokes2005findsym,campbell2006isodisplace}. The bonding characteristics were analyzed using the crystal orbital Hamilton population (COHP) method as implemented in the LOBSTER code\cite{maintz2016lobster}. Bader charge analysis was employed to estimate the effective atomic charges before and after hydrogen doping\cite{henkelman2006fast}. The occupied electronic manifold was analyzed using maximally localized Wannier functions (MLWFs) within the Wannier90 package\cite{mostofi2008wannier90,PhysRevB.65.035109}. The disentanglement energy window was set from E$_F$+(-7.386 eV) to E$_F$+(2.164 eV) , where E$_F$=5.436eV, covering all relevant bands near the Fermi level. A total of 112 Wannier functions were constructed based on Ni 3d and O 2p orbitals, including five 3d orbitals(d$_{xy}$, d$_{yz}$, d$_{xz}$, d$_{x^2-y^2}$, d$_{z^2}$) for each of the eight Ni atoms and three 2p orbitals (p$_x$, p$_y$, p$_z$) for each of the 24 O atoms. Since only the spin-up states exhibit significant changes near the Fermi level upon hydrogenation, MLWFs were constructed for the spin-up channel only. As shown in Fig.~\ref{FigS1}, the corresponding Wannier-interpolated band structure agrees well with the DFT results, demonstrating the accuracy of the constructed Wannier functions.

Based on the machine-learning interatomic potential CHGNet\cite{deng2023chgnet}, the minimum energy paths and migration barriers were calculated using the climbing-image nudged elastic band (CI-NEB) method\cite{henkelman2000climbing}, as implemented in the Atomic Simulation Environment (ASE)\cite{hjorth2017atomic}. Intermediate images were generated by linear interpolation between the initial and final states, and structural relaxations were performed using the BFGS optimizer until the maximum residual force on each atom was less than 0.05 eV/\AA. Molecular dynamics (MD) simulations based on the CHGNet potential were further carried out to investigate proton diffusion in nickelate perovskite systems. The simulations were performed in the NVT ensemble at 900 K with a time step of 0.5 fs for a total simulation time of 40 ps. It should be emphasized that CHGNet is used here to explore the qualitative influence of different chemical environments, including A-site and hydrogenation-induced lattice distortions, on proton transport in nickelate perovskites.

\section{Results and Discussion}
\subsection*{A. Structural properties of NdNiO$_3$ and HNdNiO$_3$}

In nickelate systems, the electronic structure is highly sensitive to magnetic ordering\cite{varignon2017complete,PhysRevLett.103.156401}; therefore, we first examine the magnetic ground state of NdNiO$_3$ in its high-temperature orthorhombic phase. Since NdNiO$_3$ exhibits a complex magnetic configuration characterized by the propagation vector Q=($\tfrac{1}{2}$,0,$\tfrac{1}{2}$) only at low temperatures ($T < 200 K$)\cite{PhysRevB.57.456,PhysRevB.64.144417}, we consider several conventional magnetic configurations for the high-temperature phase, including ferromagnetic (FM) and three types of antiferromagnetic orderings: A-type (A-AFM), C-type (C-AFM), and G-type (G-AFM), corresponding to interlayer, intrachain, and fully staggered antiferromagnetic arrangements, respectively. Our calculations show that orthorhombic NdNiO$_3$ favors a FM ground state, while all AFM configurations are at least 100 meV higher in energy per 40-atom supercell (Fig.~\ref{FigS2}). This result is consistent with previous DFT+U studies\cite{yoo2018metal,varignon2017complete,PhysRevB.109.205124}. Based on the orthorhombic ferromagnetic ground state, Table~\ref{Tab1} summarizes the structural parameters of NdNiO$_3$ obtained from our calculations, together with available experimental data and previous theoretical results. We obtain lattice constants of a=5.331~\AA, b=5.331~\AA, c=7.605~\AA, all of which deviate by less than 1\% from the experimental values\cite{catalano2018rare,middey2016physics}. Due to the relatively small ionic radius of the A-site rare-earth element Nd, the ideal cubic perovskite condition d$_{A-O}$=$\sqrt(2)$d$_{B-O}$ (CaTiO$_3$-type)\cite{medarde1997structural} is not satisfied. As a result, the NiO$_6$ octahedra undergo tilting to fill the extra interstitial space, leading to a deviation of the Ni-O-Ni bond angles from $180^\circ$. Specifically, two symmetry-inequivalent oxygen Wyckoff sites (O$_{8d}$, O$_{4c}$) yield Ni-O-Ni bond angles of $156.54^\circ$ and $157.62^\circ$, respectively. Consequently, the bond-length mismatch is relieved through a structural distortion toward a lower-symmetry orthorhombic phase of the GdFeO$_3$-type. This phase is characterized by two distinct octahedral rotation patterns, denoted as a$^-$a$^-$c$^0$ and a$^0$a$^0$c$^+$ in Glazer's notation\cite{glazer1972classification}, corresponding to out-of-phase rotations of neighboring octahedra along the a and b axes and in-phase rotations along the c axis, respectively (Fig.~\ref{Fig1}).

Since hydrogen is inserted at the apical oxygen site along the c-axis chain (Fig.~\ref{Fig1}), the resulting O-H bond lies within the Nd-O plane in the ab directions. The Coulomb repulsion between the proton and neighboring Nd cations predominantly leads to an expansion of the nearest-neighbor Nd-Nd distances within the ab plane. Consequently, upon hydrogenation, the lattice constants along the a and b directions increase by 6.2\%, while the c-axis remains nearly unchanged, leading to an overall volume expansion of 12.4\%, in good agreement with the experimental value of 13.2\%\cite{gao2023unveiling}. The proton bound to the oxygen site further induces antiferrodistortive octahedral rotations along the NiO$_6$ chains\cite{n8dz-gwfj}, consistent with polaron formation\cite{goodenough1986bond}, as will be demonstrated in the following section. As a result, the octahedral tilting is significantly enhanced upon hydrogenation, with the Ni-O$_{8d}$-Ni and Ni-O$_{4c}$-Ni bond angles decreasing to $142.14^\circ$ and $145.15^\circ$, respectively. And the amplitudes of the a$^-$a$^-$c$^0$ and a$^0$a$^0$c$^+$ rotation modes are also markedly increased (Fig.~\ref{Fig1}). Owing to the combined effect of increased octahedral tilting (further deviation of the Ni-O-Ni angles from 180$^\circ$) and lattice expansion in the ab plane, the average Ni-O$_{8d}$ bond length increases from 1.925~\AA\ to 2.116~\AA\ (a 9.9\% increase). In contrast, because the lattice parameter along the c direction remains nearly unchanged, the Ni-O$_{4c}$ bond length shows a much smaller increase of only 2.8\%. These relative changes in bond lengths and bond angles are consistent with previous reports\cite{gao2023unveiling}. In addition, due to the out-of-phase octahedral rotation pattern of the c-axis chain, adjacent Nd atoms stacked along the c-axis are slightly displaced relative to each other (Fig.~\ref{Fig1}, left). Hydrogenation enhances this antiferrodistortive octahedral rotation, leading to a more pronounced separation between neighboring Nd atoms (Fig.~\ref{Fig1}, right). As a consequence, the structure develops a weak monoclinic distortion, with the $\gamma$ angle decreasing from $89.85^\circ$ to $85.33^\circ$.

\begin{figure}[htbp]
\centering
\includegraphics[scale=0.52]{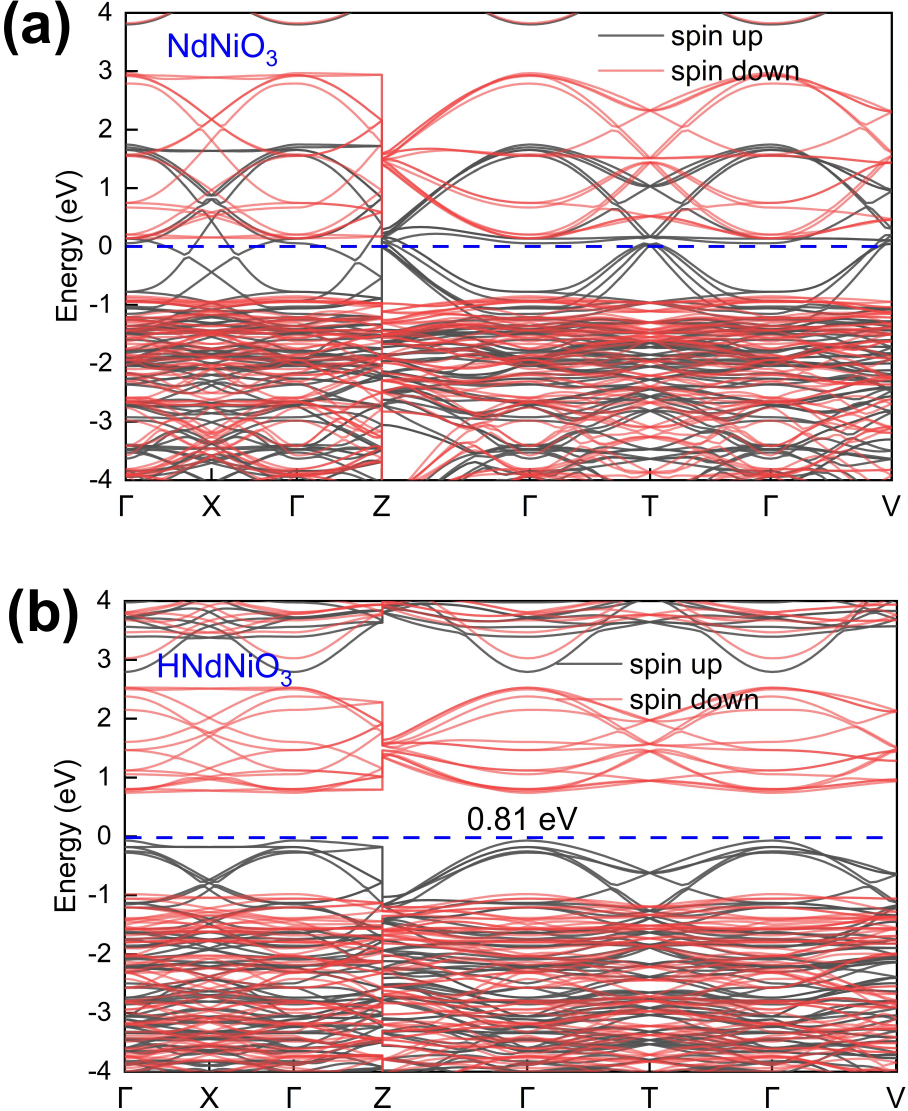}
\caption{Calculated band structure of NdNiO$_3$ (a) and HNdNiO$_3$(b).}
\label{Fig2}
\end{figure}

\begin{figure*}[htbp]
\centering
\includegraphics[scale=0.7]{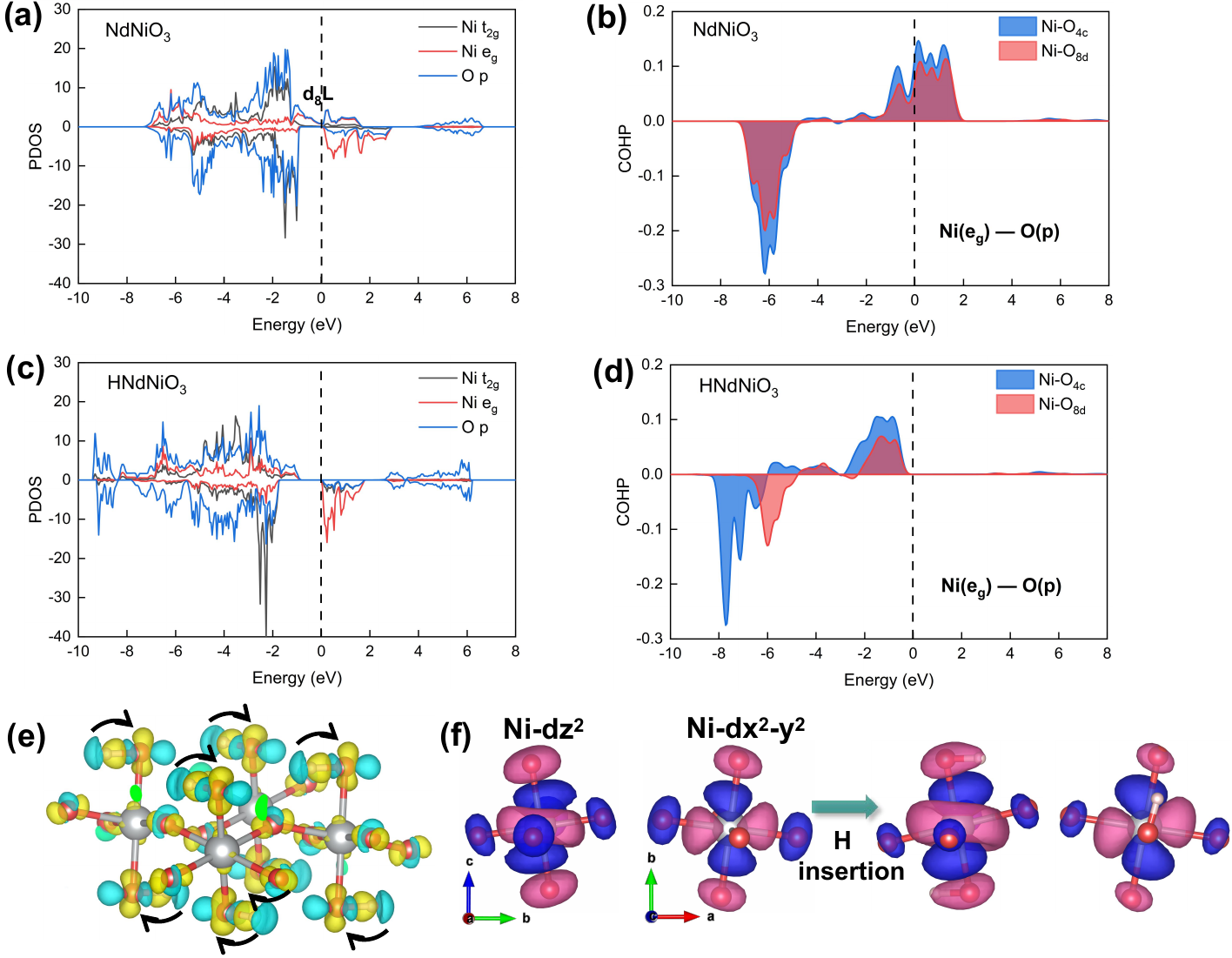}
\caption{PDOS of NdNiO$_3$ (a) and HNdNiO$_3$ (c). COHP for Ni e$_g$ - O p in NdNiO$_3$ (b) and HNdNiO$_3$ (d). (e) Charge density difference of HNdNiO$_3$, where blue and yellow regions represent charge depletion and accumulation. The isosurface value is set to 0.01 e $\AA^{-3}$. (f) Ni-centered d$_{z^2}$-like and d$_{x^2-y^2}$-like maximally localized Wannier functions (MLWFs) of NdNiO$_3$ before and after hydrogenation. The isosurface value is set to 0.7, and only the spin-up MLWFs are shown.}
\label{Fig3}
\end{figure*}

\subsection*{B. Hydrogen-induced MIT transition mechanism}
\begin{figure}[htbp]
\centering
\includegraphics[scale=0.5]{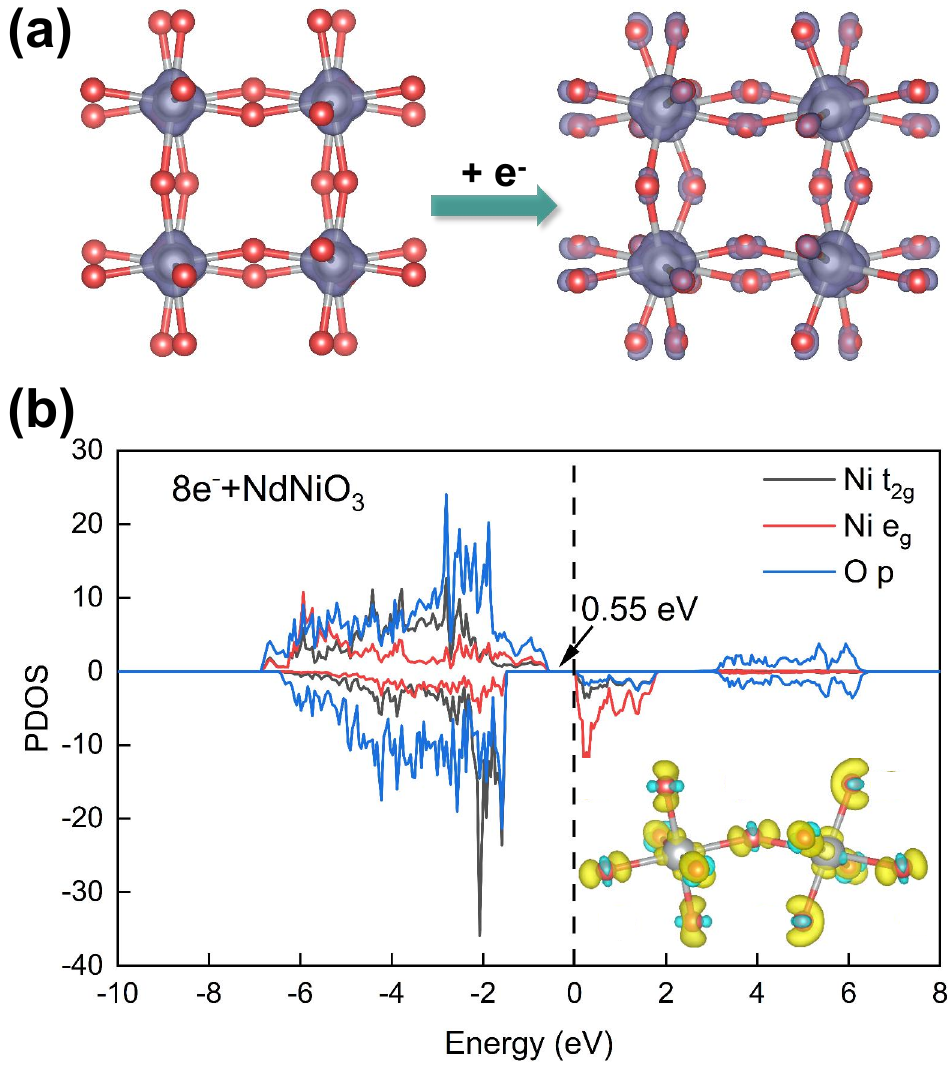}
\caption{(a) Spin density distribution of NdNiO$_3$ before and after electron doping. (b) Projected density of states (PDOS) of NdNiO$_3$ with eight electrons. The inset shows the corresponding charge density difference.}
\label{Fig4}
\end{figure}

Building on the hydrogenation-induced lattice distortions discussed above, we now examine the impact of hydrogenation on the electronic structure of NdNiO$_3$. As shown in Fig.~\ref{Fig2}, hydrogenation drives a MIT transition, consistent with previous reports on rare-earth nickelates\cite{zhou2016strongly,shi2014colossal,chen2020electron,yoo2018metal,lan2020first,gao2023unveiling} .The resulting band gap is 0.81~eV, in good agreement with prior PBEsol+U calculations\cite{yoo2018metal}. Specifically, the spin-up bands near the Fermi level shift downward below E$_F$, leading to the opening of a gap between the spin-up and spin-down channels. To further elucidate the mechanism of the MIT transition, we calculate the projected density of states (PDOS) and COHP before and after hydrogenation (Fig.~\ref{Fig3}). Since hydrogenation primarily modifies the electronic states associated with the Ni e$_g$ orbitals and O p orbitals, we focus on the directional $\sigma$-bonding arising from Ni e$_g$-O p hybridization (Fig.~\ref{Fig3}(b)). From PDOS of NdNiO$_3$ (Fig.~\ref{Fig3}(a)), the Ni e$_g$ states in the spin-up channel are distributed both below and above the Fermi level, while those in the spin-down channel are mainly located in the unoccupied region. In contrast, the Ni t$_{2g}$ states in both spin channels are fully occupied. These features indicate that Ni is in a low-spin state. This spin polarization and resulting magnetic moment in NdNiO$_3$ are primarily determined by the occupancy of the spin-up states near the Fermi level. These states mainly arise from hybridization between Ni e$_g$ and O p orbitals (Fig.~\ref{Fig3}(a)), corresponds to Ni-O antibonding character (COHP $>$ 0)(Fig.~\ref{Fig3}(b)). Moreover, the O 2p states carry a larger weight in the unoccupied part , indicating a d$_8$L electronic configuration (where L denotes an O 2p ligand hole). This is consistent with previous studies showing that the ground state of rare-earth nickelates is dominated by covalent Ni-O bonding with a d$_8$L character\cite{PhysRevB.52.15823,medarde1997structural,zhou2016strongly}. 

Upon hydrogenation, the unoccupied Ni-O hybrid states above the Fermi level become filled (Fig.~\ref{Fig3}(c)), and the corresponding antibonding features in the COHP shift below E$_F$(Fig.~\ref{Fig3}(d)). This filling of the hybrid bands is consistent with experimental X-ray absorption spectroscopy results\cite{zhou2016strongly,gao2023unveiling}. Since these antibonding states are characterized by d$_8$L, the added electrons preferentially fill these O 2p ligand hole states. Charge difference density analysis (Fig.~\ref{Fig3}(e)) further reveals a depletion of charge density around the H atom (blue) and an accumulation along the Ni-O bonding direction (yellow), indicating charge transfer from H to O. The electron filling of these O 2p holes agrees with the observations of Lei Gao et al.\cite{gao2023unveiling}. Although this band filling reduces the number of hole carriers and thus affects the electronic transport\cite{allanccon1994influence,kim2010defect,jeon2012oxygen}, the PDOS shows significant changes in the intrinsic bands of the system, which is a key factor in opening the band gap. We therefore proceed to analyze the electronic states around Ni.

Our calculations show that in pristine NdNiO$_3$, the average magnetic moment of Ni is 0.9 $\mu$B, which increases to 1.61$\mu$B upon full hydrogenation. Meanwhile, the average effective charge of Ni increases from 8.66 e$^-$ to 8.92 e$^-$. According to the electron filling scheme, the nearly twofold increase in the magnetic moment in this low-spin system indicates a transition of the Ni valence electronic configuration from d7:t$_{2g}$$^6$e$_g$$^1$ with singly occupied e$_g$ orbitals to d8:t$_{2g}$$^6$e$_g$$^2$ with half-filled e$_g$ orbitals. However, our previous analysis indicates that the electrons introduced by hydrogen doping predominantly occupy the O 2p ligand hole states. This raises a key question: where do the additional e$_g$ electrons originate from? To address this issue, we computed the maximally localized Wannier functions (MLWFs) before and after hydrogenation. As shown in Fig.~\ref{Fig3}(f), the two Ni-centered eg-like MLWFs (d$_{z^2}$-like and d$_{x^2-y^2}$-like) exhibit a clear evolution upon hydrogen insertion. Specifically, the weight of these MLWFs on the surrounding oxygen sites is reduced, and their shapes become closer to the more ideal d$_{z^2}$ and d$_{x^2-y^2}$ orbital characters. This reflects a weakening of the hybridization between Ni and O in HNdNiO$_3$, resulting in more localized Ni e$_g$ electrons and the evolution of the Ni electronic configuration from d7\cite{PhysRevLett.112.106404} to d8. 

The above results demonstrate both the filling of the O 2p ligand hole states and the localization of the Ni e$_g$ electrons. To clarify the connection between these two phenomena, we further computed NdNiO$_3$ with only eight extra electrons to isolate the effect of electron filling. As shown in the PDOS in Fig.~\ref{Fig4}(b), the injected electrons predominantly occupy the d$_8$L Ni-O antibonding states associated with O 2p ligand holes. The charge density difference for both the electron-doped (inset of Fig.~\ref{Fig4}(b)) and hydrogen-doped (Fig.~\ref{Fig3}(e)) systems consistently reveals pronounced localization of charge density along the Ni-O bonding direction. Simultaneously, significant elongations of the Ni-O bonds are also observed in electron-doped symstem, with the Ni-O$_{8d}$ and Ni-O$_{4c}$ bond lengths increasing to 2.213 and 2.045 \AA, respectively. This electron filling of the O 2p ligand hole states, accompanied by lattice distortion, is also evidenced by the spin-density redistribution and the pronounced structural distortion shown in Fig.~\ref{Fig4}(a). These results indicate the formation of electron polarons in the antibonding states of Ni-O hybridization, which locally distort the lattice, elongate the Ni-O bonds, and weaken the Ni-O hybridization. The reduced hybridization drives the localization of Ni d electrons, giving rise to a half-filled e$_g$ configuration with strong on-site Coulomb repulsion and ultimately leading to a Mott transition. Furthermore, the electron-doped system exhibits a band gap of 0.55 eV, which is smaller than the 0.81 eV gap obtained for HNdNiO$_3$. This result indicates that the protons also act as polarizing centers, weakening the Ni-O hybridization and thereby further promoting the localization of Ni e$_g$ electrons. The combined effect of electron and proton polarons on the elongation of the Ni-O bonds is summarized in Table~\ref{Tab1}. In a word, the filling of the O 2p ligand hole states in the d$_8$L hybridized configuration promotes polaron formation, which drives the originally itinerant Ni e$_g$ electrons toward localization. This lattice-distortion-mediated electronic transition mechanism is consistent with the picture proposed by Chen\cite{chen2019revealing}.

\begin{figure}[htbp]
\centering
\includegraphics[scale=0.35]{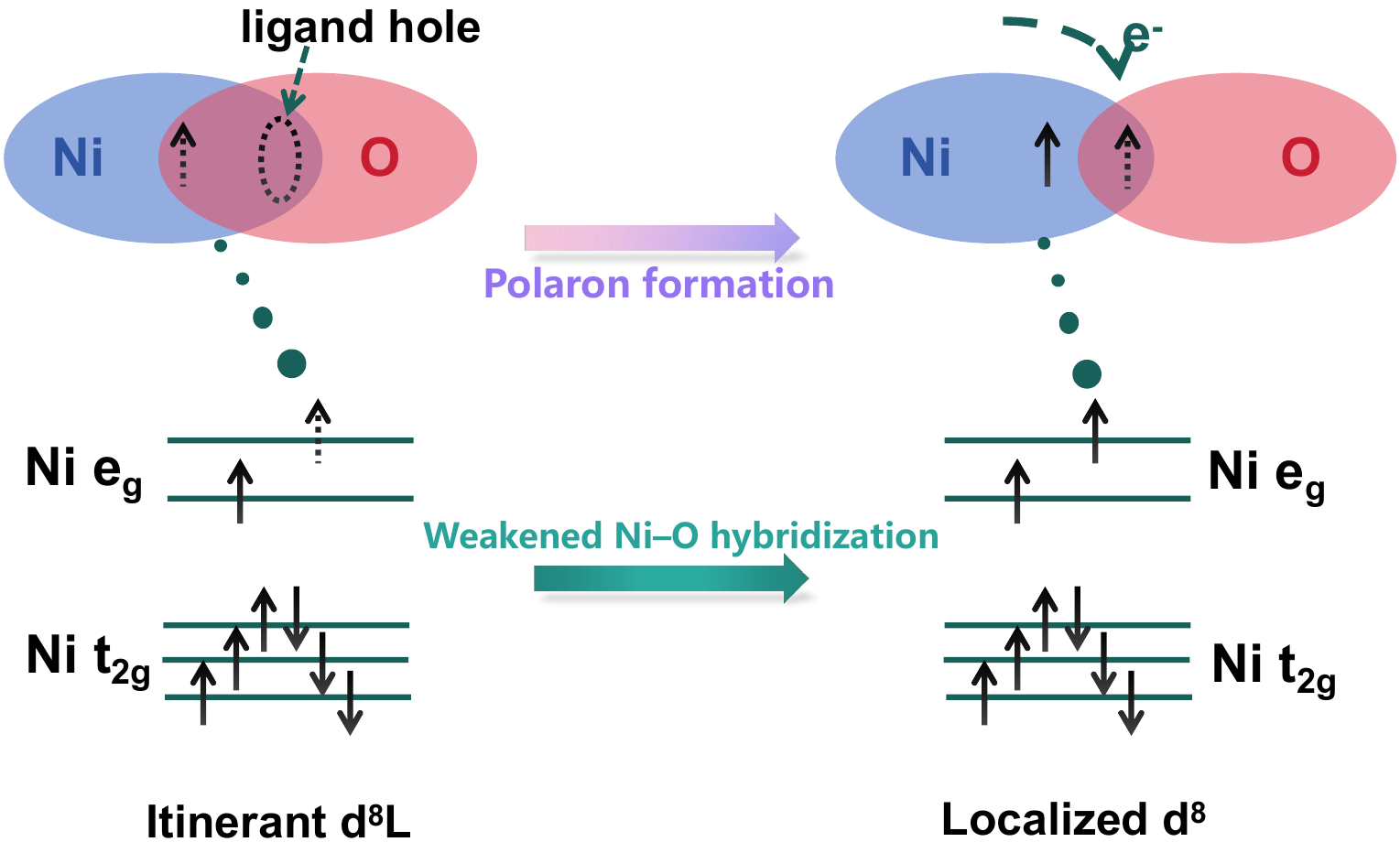}
\caption{Schematic illustration of the hydrogen-induced MIT transition mechanism in ReNiO$_3$.}
\label{Fig5}
\end{figure}

Based on the above analysis of the electronic structures of O and Ni, we propose that the MIT transition in hydrogenated rare-earth nickelates proceeds as follows (Fig.~\ref{Fig5}). The pristine system is a p-type degenerate semiconductor\cite{kim2010defect,jeon2012oxygen}, where transport is dominated by hole carriers. Upon hydrogenation, the doped electrons preferentially occupy the O 2p ligand hole states associated with the hybridized d$_8$L configuration, promoting electron-polaron formation. The resulting electron polarons, together with proton polarons, weaken the Ni-O $\sigma$ hybridization and thereby drive the originally itinerant Ni e$_g$ electrons toward localization, leading to the formation of a local d8(t$_{2g}$$^6$e$_g$$^2$) configuration. The half-filled e$_g$ manifold then experiences strong on-site Coulomb repulsion, resulting in the opening of a Mott gap. This mechanism naturally explains the site and orbital selective Mott transition reported by Bhat et al.\cite{bhat2024dynamical}. Within our physical picture, polaron formation serves as the key bridge linking ligand hole filling and Ni e$_g$ electron localization, with these two processes acting jointly to drive the system into the insulating state.

\subsection*{C. Proton conduction in ReNiO$_3$ (Re = Nd, Sm)}

\begin{figure}[htbp]
\centering
\includegraphics[scale=0.45]{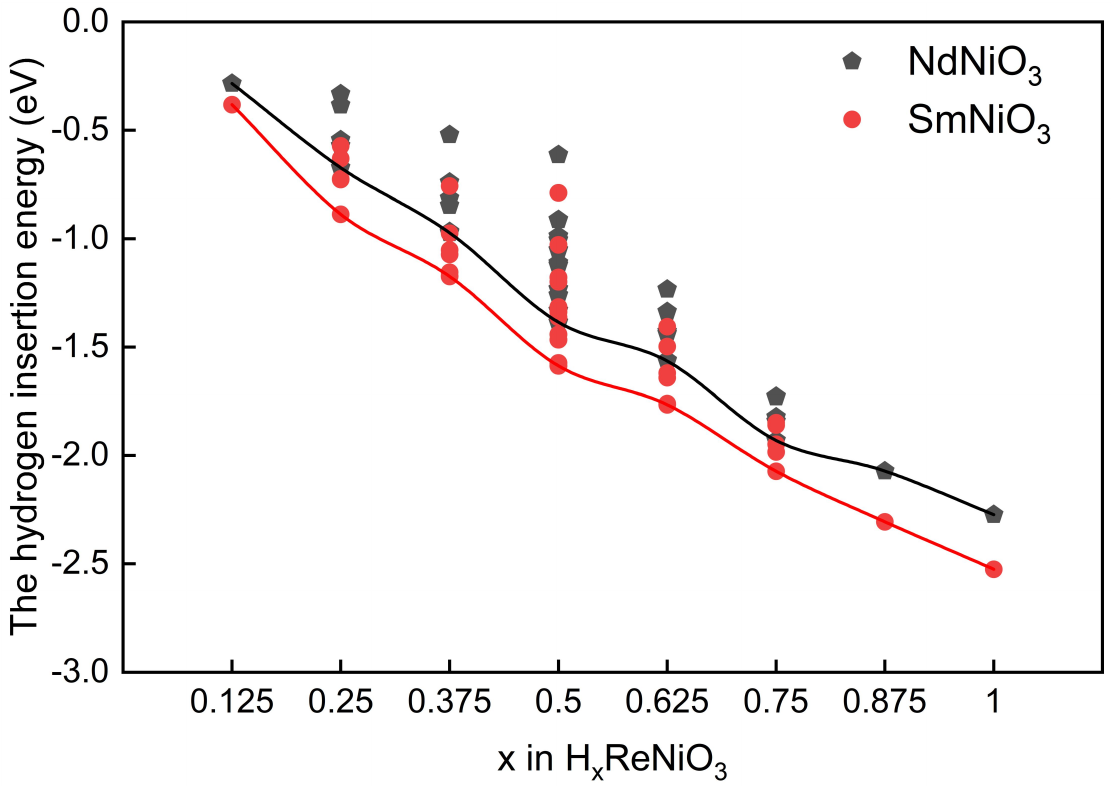}
\caption{Hydrogen insertion energy as a function of hydrogen concentration x in H$_x$ReNiO$_3$ (Re = Nd, Sm).}
\label{Fig6}
\end{figure}

\begin{figure*}[htbp]
\centering
\includegraphics[scale=0.53]{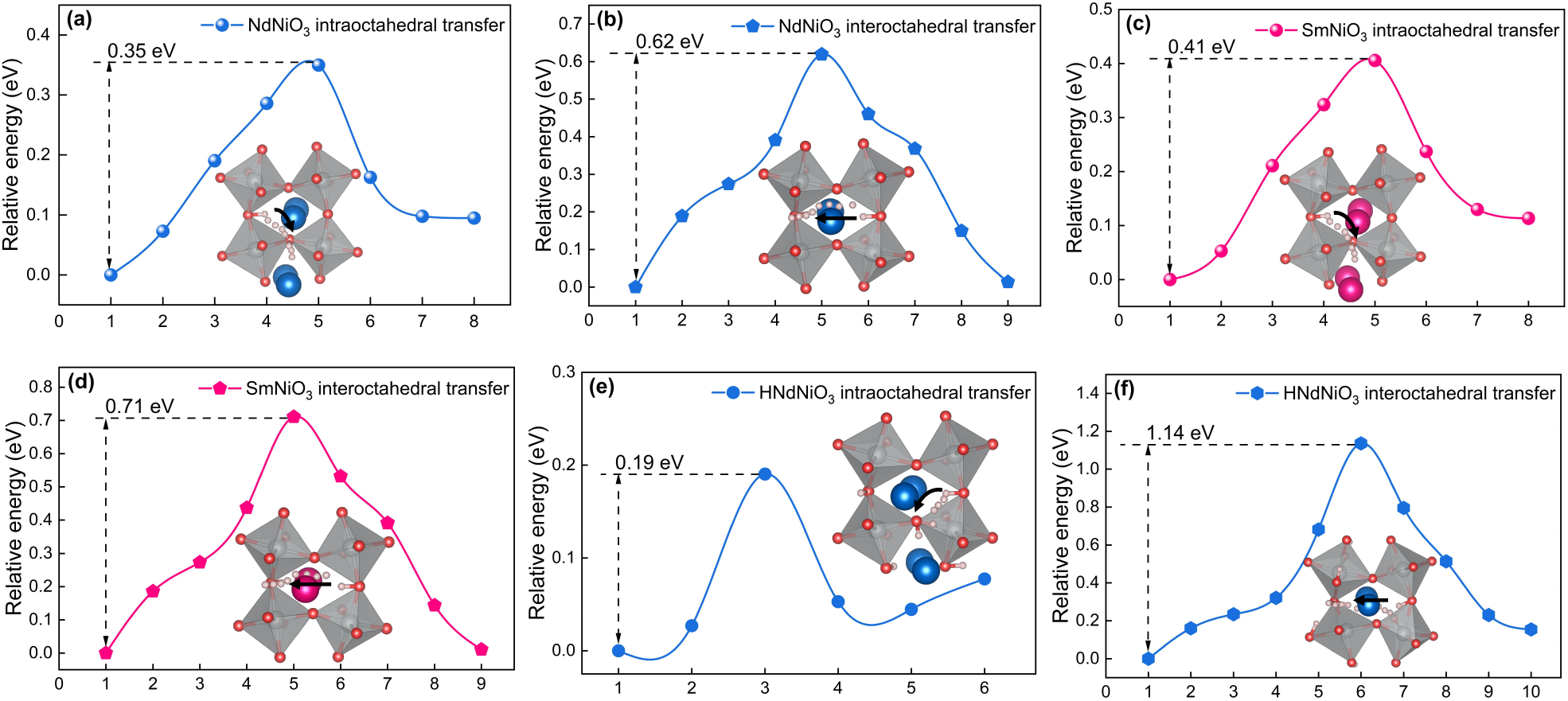}
\caption{Energy barriers and schematic pathways for intraoctahedral (a) and interoctahedral (b) transfer in NdNiO$_3$, and for intraoctahedral (c) and interoctahedral (d) transfer in SmNiO$_3$. Panels (e) and (f) show the corresponding intra and interoctahedral transfer processes in hydrogenated NdNiO$_3$. Arrows indicate the transfer directions.}
\label{Fig7}
\end{figure*}

\begin{figure}[htbp]
\centering
\includegraphics[scale=0.37]{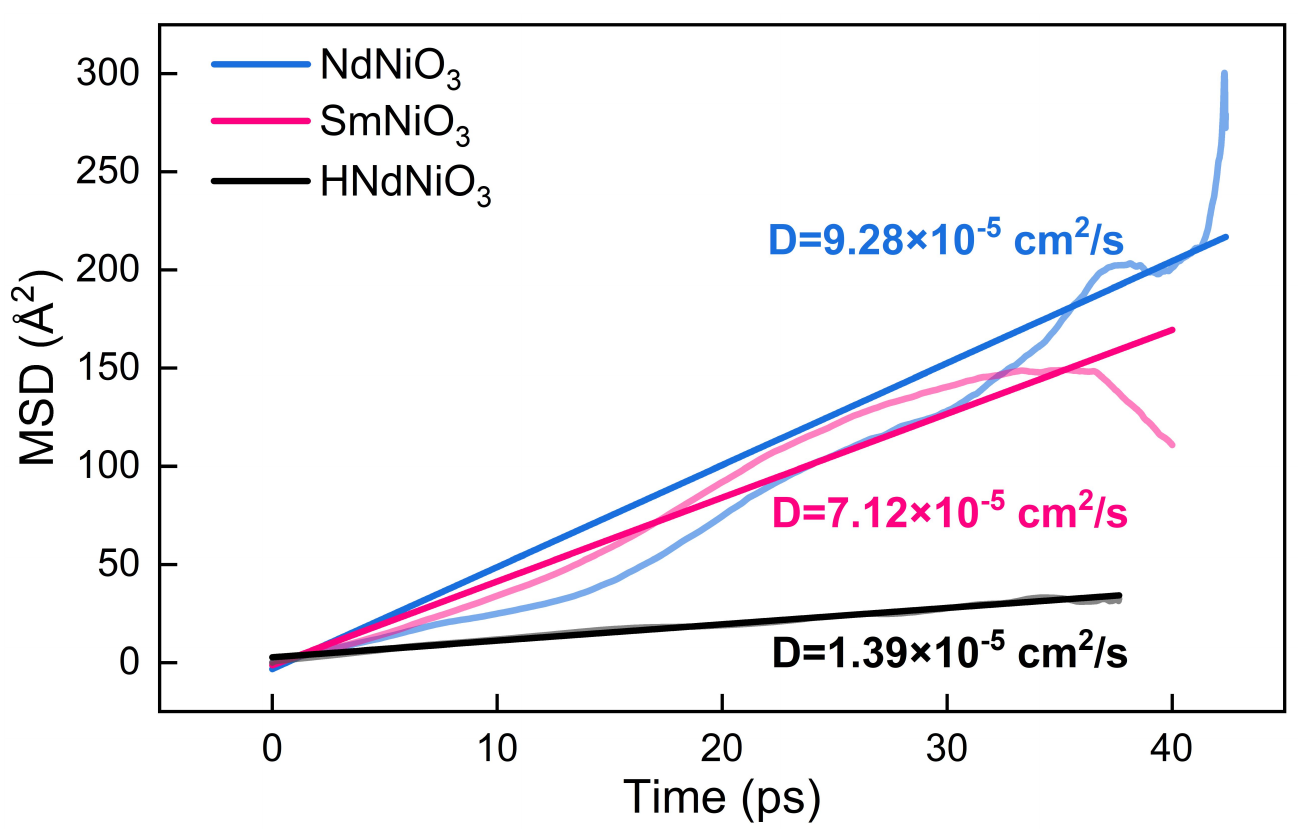}
\caption{Mean-square displacement (MSD) extracted from $\sim 40$ ps proton trajectories in NdNiO$_3$, SmNiO$_3$, and HNdNiO$_3$. The diffusion coefficients D are obtained from linear fits to the MSD-t curves based on MSD(t)=2dDt (where d is the dimensionality). The shaded curves represent the calculated MSD as a function of time, while the solid lines denote the corresponding linear fits.}
\label{Fig9}
\end{figure}

While the hydrogen-induced insulating state enables nickelates to function as proton-conducting SOFC electrolytes, their practical performance is ultimately determined by proton transport. It is governed by both the proton concentration and the proton diffusion capability\cite{PhysRevB.82.014103}. In this section, we quantify these two aspects by calculating the hydrogen insertion energy and the proton migration barrier to evaluate the proton conduction potential of NdNiO$_3$. In order to elucidate the differences in proton conduction among rare-earth nickelates, we also consider SmNiO$_3$, which has been extensively studied both experimentally and theoretically. We calculate the hydrogen insertion energy by introducing 1-8 H atoms into ReNiO$_3$ (Re = Nd, Sm) supercells containing 8 Re, 8 Ni, and 24 O atoms, corresponding to the reaction energies $\Delta H(x)$ of H$_x$ReNiO$_3$ (x=0.125, 0.25, 0.375, 0.5, 0.625, 0.75, 0.875, 1). It is defined as:
\begin{equation}
\Delta H(x) = E(\mathrm{H}_x\mathrm{ReNiO}_3) - E(\mathrm{ReNiO}_3) - \frac{x}{2}E(\mathrm{H}_2)
\end{equation}

where $E(\mathrm{H}_x\mathrm{ReNiO}_3)$ and $E(\mathrm{ReNiO}_3)$ are the total energies of the supercell after and before hydrogen insertion, respectively, and $E(\mathrm{H}_2)$ is the DFT-calculated energy of an isolated H$_2$ molecule. The hydrogen insertion energies in both systems are negative (Fig.~\ref{Fig6}), indicating that hydrogen uptake in ReNiO$_3$ is a spontaneous exothermic process. Moreover, the insertion energy decreases with increasing hydrogen concentration in H$_x$ReNiO$_3$, suggesting that the system thermodynamically favors higher hydrogen content. These behaviors can be understood from the COHP analysis (Fig.~\ref{Fig3}(b),(d)). In pristine ReNiO$_3$, antibonding states cross the Fermi level, indicating the instability of electrons near E$_F$. Upon full hydrogenation (x=1), these antibonding states become fully occupied and shift below the Fermi level, stabilizing the system\cite{maintz2016lobster}. This provides a microscopic explanation for previous findings that strongly correlated perovskites tend to spontaneously incorporate hydrogen for charge compensation due to the multivalence of B-site cations\cite{islam2020computational,linghu2025multivalent}. In addition, over the entire hydrogen concentration range, the hydrogen insertion energies of SmNiO$_3$ are more negative than those of NdNiO$_3$, indicating a stronger tendency for hydrogen incorporation, consistent with experimental observations\cite{chen2015self}. This can be attributed to the smaller ionic radius of Sm, which leads to a reduced tolerance factor and thus a less stable lattice. Indeed, the formation energy of SmNiO$_3$ (-4.143~eV) is higher than that of NdNiO$_3$ (-4.436~eV), implying a larger thermodynamic driving force for stabilization via hydrogen insertion. Furthermore, as discussed below, the longer Ni-O bond length in SmNiO$_3$ weakens the Ni-O covalency, resulting in a greater electron density around oxygen sites, which further facilitates proton incorporation.

In oxides, protons are bound to oxygen sites. For each Wyckoff oxygen site, four mutually perpendicular interstitial orientations are available for the proton (Fig.~\ref{FigS3}). We select the lowest-energy interstitial site as the proton orientation at each oxygen site, and compute the migration barriers for proton transfer between apical and equatorial oxygen sites (O$_{4c}$-O$_{8d}$) and between neighboring apical sites (O$_{4c}$-O$_{4c}$). These two processes correspond primarily to intraoctahedral and interoctahedral transfer, respectively, and constitute the essential pathways through which proton transport percolates throughout the ReNiO$_3$ lattice. The rotation process is inherently included in the calculated pathways and serves to reorient the proton toward stable interstitial sites, without introducing an additional energy barrier. 

As shown in Fig.~\ref{Fig7}(a),(b), the transfer barriers for intraoctahedral and interoctahedral transfer in NdNiO$_3$ are 0.35 and 0.62~eV, respectively. The higher interoctahedral transfer barrier can be attributed to the closer proximity of the proton to the A-site cation (Fig.~\ref{Fig8}(a)). Compared to NdNiO$_3$, SmNiO$_3$ exhibits higher transfer barriers for both pathways, increasing to 0.41 and 0.71 eV, respectively (Fig.~\ref{Fig7}(c)(d)), due to the enhanced Coulomb repulsion from the smaller A-site cations and the stronger proton-oxygen binding (Fig.~\ref{Fig8},Table.~\ref{TabS1}). This is further supported by the proton diffusion coefficients at 900~K extracted from MD simulations (Fig.~\ref{Fig9}), yielding $D=9.28\times10^{-5} cm^2/s$ for NdNiO$_3$ and $7.12\times10^{-5} cm^2/s$ for SmNiO$_3$. Hydrogenation significantly reduces the intraoctahedral transfer barrier (Fig.~\ref{Fig7}(e), 0.19 eV), while markedly increasing the interoctahedral barrier (Fig.~\ref{Fig7}(f), 1.14 eV) compared to pristine NdNiO$_3$, owing to the competition between hydrogen-induced lattice expansion and enhanced octahedral distortions (Fig.~\ref{FigS4}). These results indicate that hydrogenation facilitates proton transport in ReNiO$_3$ along the (001) direction via intraoctahedral transfer, consistent with the one-dimensional percolation pathway extracted using the SoftBV program\cite{chen2019softbv} (Fig.~\ref{FigS4}(b)), thereby facilitating proton transfer across the electrode-electrolyte interface\cite{zhou2016strongly}. However, it is unfavorable for three-dimensional proton transport within the electrolyte layer. As shown in Fig.~\ref{Fig9}, the proton diffusion coefficient of HNdNiO$_3$ is $1.39\times10^{-5} cm^2/s$, which is approximately seven times lower than that of pristine NdNiO$_3$. This reduction originates from proton trapping at low-energy sites\cite{n8dz-gwfj}, consistent with the bond-valence energy landscape showing a less interconnected low-energy diffusion network in HNdNiO$_3$ than in NdNiO$_3$ (Fig.~\ref{FigS4}(c),(d)).

\section{Summary}
Based on the orthorhombic ferromagnetic ground state of NdNiO$_3$, we find that hydrogen insertion induces lattice expansion, enhances octahedral tilting and rotational distortion modes, elongates the Ni-O bonds. More importantly, hydrogenation opens a band gap of 0.81~eV, driving a MIT transition. Through further analysis, we establish a comprehensive physical picture of the MIT transition. Upon hydrogenation, the doped electrons occupy the O 2p ligand hole states associated with the hybridized d$_8$L configuration, promoting electron-polaron formation. The resulting electron polarons, together with proton polarons, weaken the Ni-O hybridization and thereby drive the originally itinerant Ni e$_g$ electrons toward localization, leading to the formation of a local d8(t$_{2g}$$^6$e$_g$$^2$) configuration. The resulting Coulomb repulsion between neighboring half-filled e$_g$ orbitals opens a Mott gap. In addition, calculations of proton migration barriers and diffusion coefficients reveal that, compared to NdNiO$_3$, SmNiO$_3$ exhibits a stronger tendency for hydrogen incorporation, but slightly weaker proton diffusion capability. Hydrogenation facilitates proton migration through the intraoctahedral pathway, thereby enhancing one-dimensional diffusion, while impeding three-dimensional conduction. In a word, we elucidate the mechanism of the hydrogenation-induced MIT transition in rare-earth nickelates and provide guidance for experimental screening of strongly correlated oxides as electrolyte materials. Our quantitative analysis of proton conduction in rare-earth nickelates offers theoretical insight into improving their proton conductivity.

\noindent
\underbar{\bf Acknowledgements:}
This work was supported by Beijing Natural Science
Foundation (Nos.1252022 and 1242022), and National Natural Science Foundation of China (Nos. 12404463 and 12474218).

\noindent
\underbar{\bf Data availability:}
The data that support the findings of this article are openly available\cite{ma_2026_21278339}.


\appendix
\section*{Appendix}

\setcounter{equation}{0}
\setcounter{figure}{0}
\renewcommand{\theequation}{A\arabic{equation}}
\renewcommand{\thefigure}{A\arabic{figure}}
\renewcommand{\thesubsection}{A\arabic{subsection}}

\subsection*{1. Wannier-interpolated band structures}
\begin{figure}[htbp]
\centering
\includegraphics[scale=0.5]{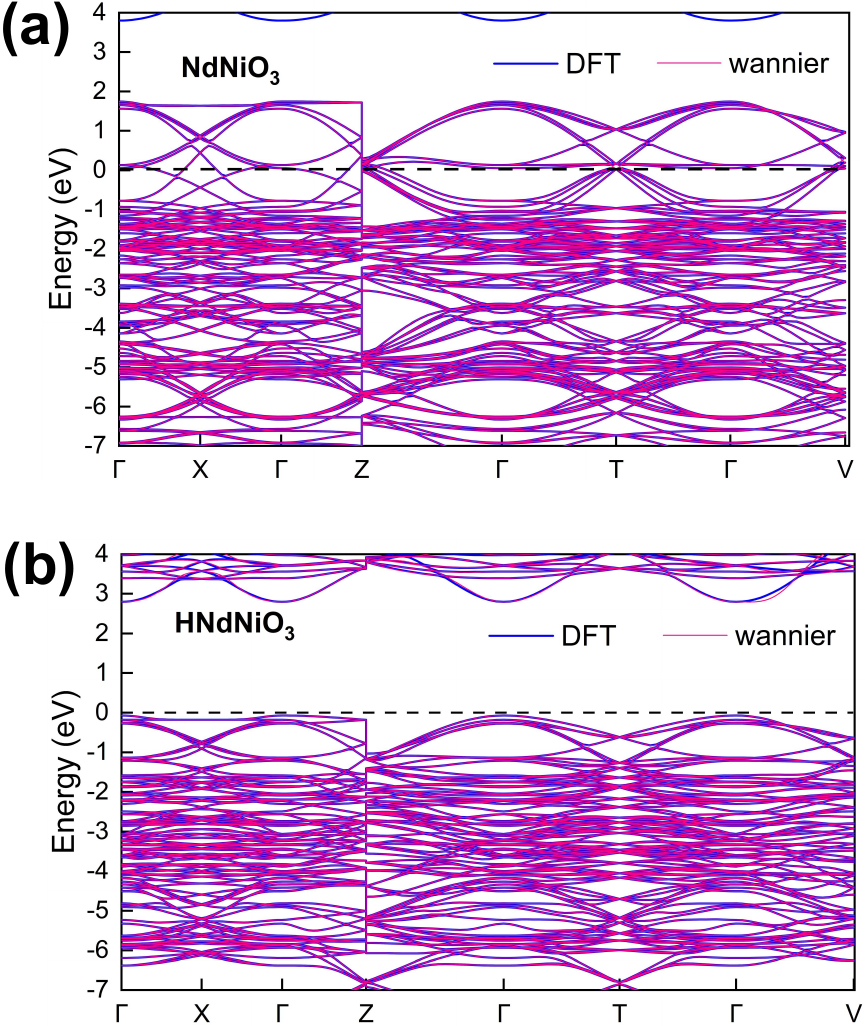}
\caption{Wannier-interpolated (pink) and DFT-calculated (blue) spin-up band structures of NdNiO$_3$ (a) and HNdNiO$_3$ (b).}
\label{FigS1}
\end{figure}

As shown in Fig.~\ref{FigS1}, the band structure obtained from the Wannier-interpolated tight-binding model for the spin-up channel agrees well with the DFT results, demonstrating the accuracy of the constructed Wannier functions.

\subsection*{2. Magnetic ground state of NdNiO$_3$}
\begin{figure}[htbp]
\centering
\includegraphics[scale=0.5]{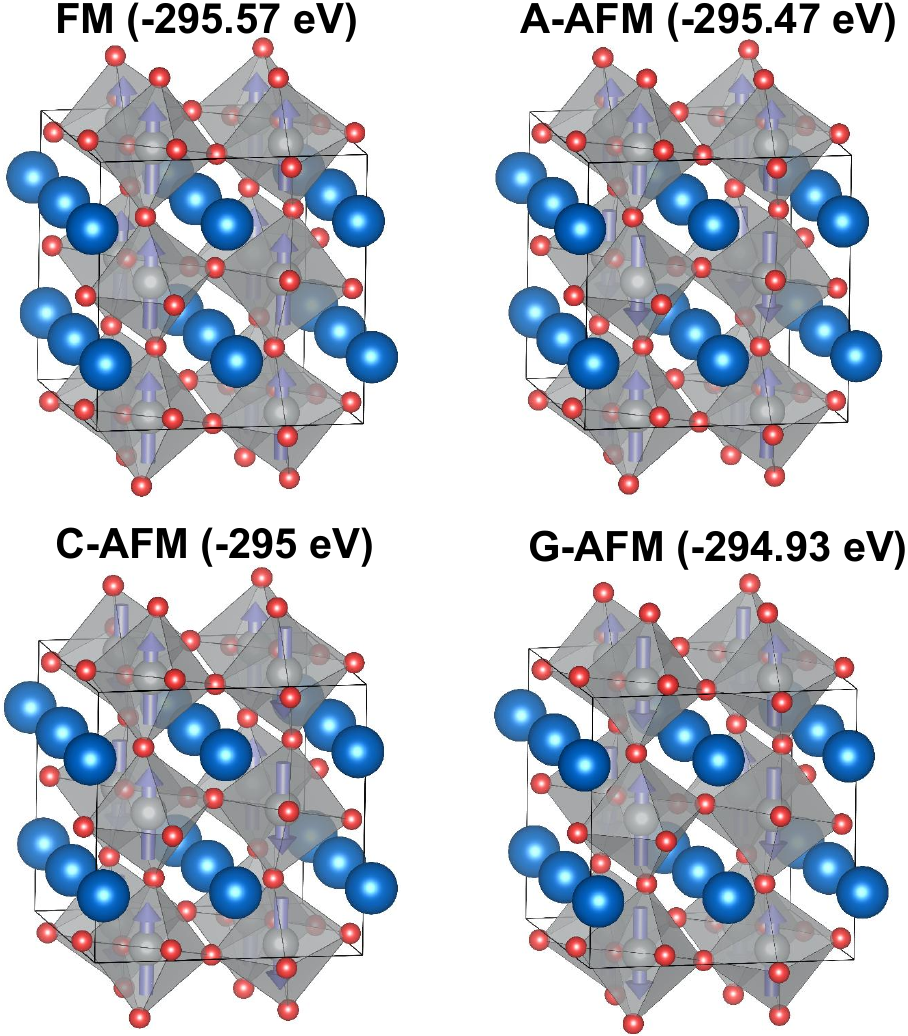}
\caption{Schematic illustrations of the FM, A-AFM, C-AFM, and G-AFM configurations.}
\label{FigS2}
\end{figure}

As shown in Fig.~\ref{FigS2}, based on a 40-atom orthorhombic NdNiO$_3$ supercell, our PBEsol+U calculations yield the relative energy ordering of different magnetic configurations as FM $<$ A-AFM $<$ C-AFM $<$ G-AFM. Therefore, the FM state is identified as the magnetic ground state of orthorhombic NdNiO$_3$.

\subsection*{3. The structural parameters of SmNiO$_3$}
\begin{table}[htbp]
\centering
\caption{Lattice constants, bond lengths, bond angles, and structural distortion modes of SmNiO$_3$}
\label{TabS1}
\setlength{\tabcolsep}{8pt}
\renewcommand{\arraystretch}{1.2}
\begin{tabular}{lc}
\toprule
& \multicolumn{1}{c}{SmNiO$_3$}  \\
\cmidrule(lr){2-2} 
& PBEsol+U \\
\midrule
$a$ (\AA) & 5.318   \\
$b$ (\AA) & 5.315   \\
$c$ (\AA) & 7.531   \\
$\alpha$ ($^\circ$) & 90  \\
$\beta$ ($^\circ$)  & 90  \\
$\gamma$ ($^\circ$) & 88.22 \\
\midrule
Ni--O$_{8d}$ (\AA) & 1.935 \\
Ni--O$_{4c}$ (\AA) & 1.93  \\
$\angle$Ni--O$_{8d}$--Ni ($^\circ$) & 152.582   \\
$\angle$Ni--O$_{4c}$--Ni ($^\circ$) & 154.714   \\
\midrule
a$^-$a$^-$c$^0$ (\AA/f.u.) & 1.2451   \\
a$^0$a$^0$c$^+$ (\AA/f.u.) & 0.9252   \\
\bottomrule
\end{tabular}
\end{table}

Table~\ref{TabS1} presents the structural parameters of SmNiO$_3$. Compared to NdNiO$_3$, the smaller A-site ionic radius in SmNiO$_3$ leads to a larger lattice mismatch between the A- and B-site, resulting in enhanced octahedral tilting and rotation. Consequently, the Ni-O-Ni bond angles are reduced, and the amplitudes of the a$^-$a$^-$c$^0$/a$^0$a$^0$c$^+$ rotation modes are increased, leading to elongated Ni-O bonds.

\begin{figure*}[htbp]
\centering
\includegraphics[scale=0.5]{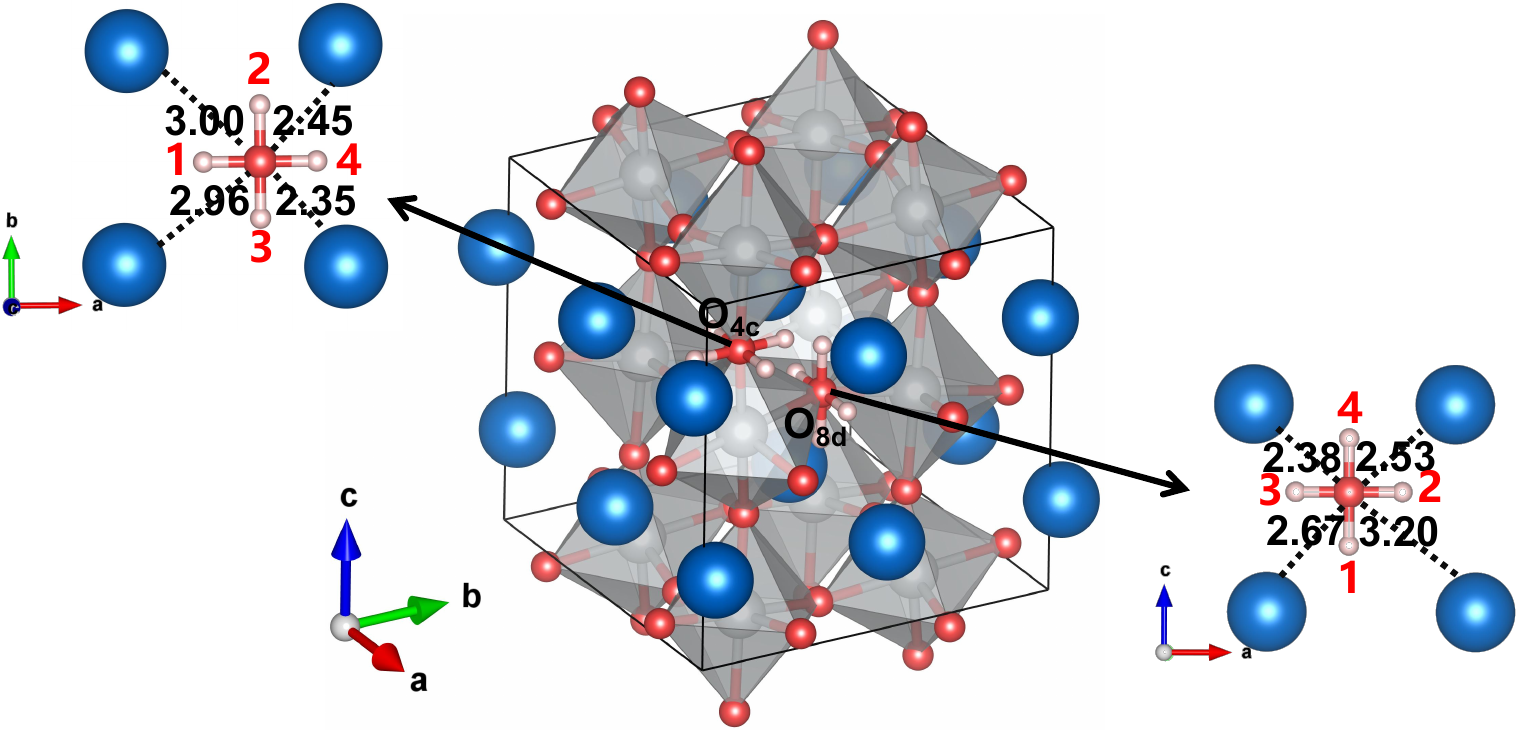}
\caption{Schematic illustration of proton interstitial orientations bound to apical O$_{4c}$ and equatorial O$_{8d}$ sites.}
\label{FigS3}
\end{figure*}

\begin{figure}[htbp]
\centering
\includegraphics[scale=0.67]{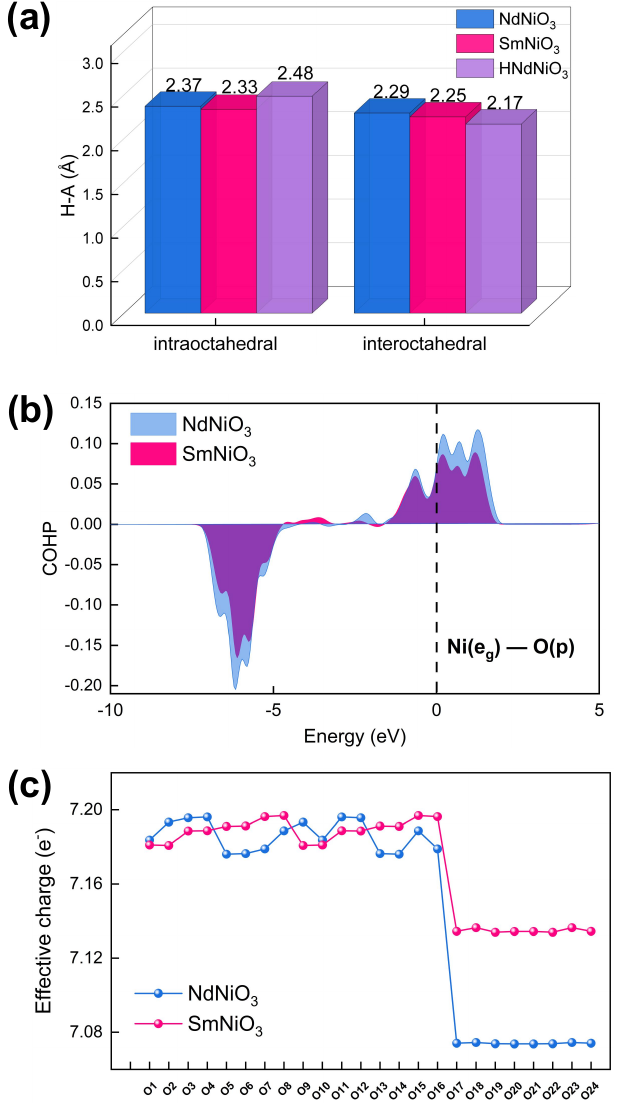}
\caption{(a) Distance between the proton at the transition state and the nearest A-site cations during intra and interoctahedral transfer in NdNiO$_3$, SmNiO$_3$, and HNdNiO$_3$. (b) COHP of the Ni e$_g$–O p hybridization ($\sigma$bonding) in NdNiO$_3$ and SmNiO$_3$. (c) Effective charges of all oxygen atoms in NdNiO$_3$ and SmNiO$_3$.}
\label{Fig8}
\end{figure}

\begin{figure*}[htbp]
\centering
\includegraphics[scale=0.5]{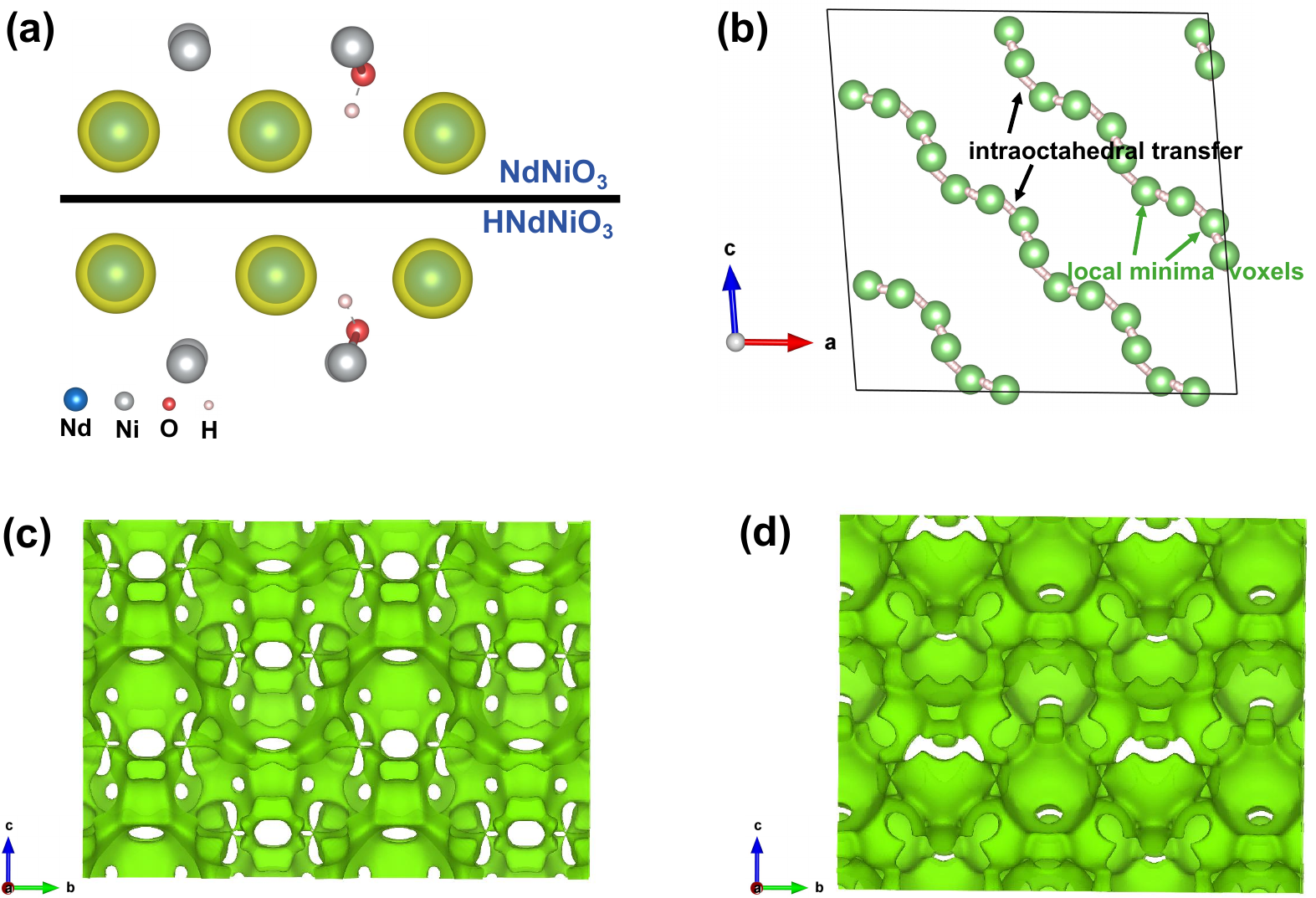}
\caption{(a). Schematic illustration of the transition state for interoctahedral proton transfer in NdNiO$_3$ and HNdNiO$_3$. Only Nd, Ni, the transferring proton, and the nearest oxygen atom are shown. The yellow shading denotes the region of positive electrostatic potential around Nd. (b). The one-dimensional percolation pathway extracted using the SoftBV program.  Bond-valence based energy landscape (BVEL) for a proton in NdNiO$_3$ (c) and HNdNiO$_3$ (d), where the green isosurface corresponds to 0.3~eV.}
\label{FigS4}
\end{figure*}

\subsection*{4.  Proton configurations at oxygen sites}

As shown in Fig.~\ref{FigS3}, protons bound to apical O$_{4c}$ and equatorial O$_{8d}$ sites exhibit four mutually perpendicular interstitial orientations. For each oxygen site, the configurational energies of these orientations depend on the distances between the proton and the two nearest A-site cations. The interstitial orientations (labeled 1–4 in Fig.~\ref{FigS3}) are ordered according to decreasing proton-A-site distance, and the corresponding relaxed energies increase accordingly. Notably, the proton cannot be stabilized at site 4, where the Coulomb repulsion is strongest.

\subsection*{5. Proton diffusion in NdNiO$_3$ and SmNiO$_3$}
The transfer barriers for intraoctahedral and interoctahedral transfer in NdNiO$_3$ are 0.35 and 0.62~eV, respectively. The higher barrier for interoctahedral transfer arises because the proton passes through the A-O plane at the transition state, resulting in a shorter distance to the A-site cations compared to the intraoctahedral process (Fig.~\ref{Fig8}(a)). This leads to stronger Coulomb repulsion and a less stable, higher-energy transition state. Proton transport in ReNiO$_3$ therefore preferentially proceeds via intraoctahedral transfer for long-range diffusion, consistent with previous reports\cite{lan2020first}. 

Compared to NdNiO$_3$, SmNiO$_3$ exhibits higher transfer barriers for both pathways, increasing to 0.41 and 0.71 eV, respectively. This is because, on the one hand, the smaller ionic radius of Sm induces lattice contraction, reducing the distance between the proton and the A-site cations at the transition state (Fig.~\ref{Fig8}(a)), thereby enhancing Coulomb repulsion; on the other hand, the smaller A-site ionic radius leads to a larger lattice mismatch between the A- and B-site, resulting in stronger octahedral tilting distortions\cite{medarde1997structural}, smaller Ni-O-Ni bond angles, and longer Ni-O bonds (Table.~\ref{TabS1}). Consequently, the Ni-O covalency is weakened (Fig.~\ref{Fig8}(b), where the COHP peaks of NdNiO$_3$ are higher than those of SmNiO$_3$ over the entire energy range), and electrons are more localized around oxygen sites (Fig.~\ref{Fig8}(c), where most oxygen sites in SmNiO$_3$ have larger effective charges). This promotes stronger O-H bond formation, thereby hindering proton transfer. 

\subsection*{6. Comparison of proton diffusion in NdNiO$_3$ and HNdNiO$_3$}
Hydrogenation significantly reduces the intraoctahedral transfer barrier, while markedly increasing the interoctahedral barrier compared to pristine NdNiO$_3$. This trend is consistent with the change in the distance between the proton and the Nd cations at the transition state (Fig.~\ref{Fig8}(a)). For the intraoctahedral transfer, the increase in the H-Nd distance originates from the lattice expansion induced by hydrogen insertion, as discussed above. In contrast, for interoctahedral transfer, the H-Nd distance decreases because the proton passes through the Nd-O plane at the transition state, and hydrogenation enhances the octahedral tilting, causing the equatorial oxygen to deviate further from the Ni-O plane, thereby bringing the proton at the transition state closer to the Nd cations (Fig.~\ref{FigS4}).

Figure~\ref{FigS4}(b) shows the one-dimensional percolation pathway of proton in HNdNiO$_3$. The green spheres represent local minima voxels extracted from the BVEL for a proton. Since the distances between adjacent minima are comparable to those between protons at neighboring oxygen sites in DFT calculations, the segments connecting these minima correspond to intra-octahedral transfer. These transfers form a periodic pathway network with pronounced directionality. In contrast, the pristine system does not exhibit such a well-defined one-dimensional pathway. Therefore, hydrogenation facilitates long-range proton transport along the (001) direction via intra-octahedral transfer, which is relevant to transport across the electrode–electrolyte interface. However, hydrogenation is unfavorable for three-dimensional proton diffusion within the electrolyte. As shown in Fig.\ref{FigS4}(c) and (d), the low-energy channels for protons in NdNiO$_3$ are more topologically interconnected throughout the system, whereas those in HNdNiO$_3$ are more localized, indicating stronger trapping of protons at local energy minima. As a result, the proton diffusion coefficient is reduced (Fig.~\ref{Fig9}).

\bibliography{ref}

\end{document}